\documentclass[reprint,showpacs,preprintnumbers,amsmath,amssymb, aip]{revtex4-1}
\usepackage{graphicx,wasysym}
\usepackage{dcolumn}
\usepackage{bm}
\usepackage{natbib}
\usepackage{tabulary}
\usepackage{natmove}

\begin{document}

\preprint{1}

\title{Rate constants in spatially inhomogeneous systems}
\author{Addison J. Schile}
\affiliation{Department of Chemistry, University of California, Berkeley}
\affiliation{Lawrence Berkeley National Laboratory, University of California, Berkeley}
\author{David T. Limmer}
 \email{dlimmer@berkeley.edu.}
\affiliation{Department of Chemistry, University of California, Berkeley}
\affiliation{Lawrence Berkeley National Laboratory, University of California, Berkeley}
\affiliation{Kavli Energy NanoSciences Institute, University of California, Berkeley}

\date{\today}
\begin{abstract}
We present a theory and accompanying importance sampling method for computing rate constants in spatially inhomogeneous systems.
Using the relationship between rate constants and path space partition functions, we illustrate that the relative change in the rate of a rare event through space is isomorphic to the calculation of a free energy difference, albeit in a trajectory ensemble. 
Like equilibrium free energies, relative rate constants can be estimated by importance sampling. An extension to transition path sampling is proposed that combines biased path ensembles and weighted histogram analysis to accomplish this estimate. 
We show that rate constants can also be decomposed into different contributions, including relative changes in stability, barrier height and flux. 
This decomposition provides a means of interpretation and insight into rare processes in complex environments.  
We verify these ideas with a simple model of diffusion with spatially varying diffusivity and illustrate their utility in a model of ion pair dissociation near an electrochemical interface.
\end{abstract}

\maketitle

Macroscopic rate constants quantify the characteristic timescale for an ensemble of components to transition between two stable states. These states may be the reactants and products of a chemical transformation or the collective reorganizations accompanying a phase transformation. Microscopically, rate constants are related to dynamical fluctuations as codified in time-correlation functions of populations of the stable states\cite{chandler1978statistical}.  In the presence of spatial inhomogenieties, such as extended interfaces, dynamical fluctuations need not be the same throughout space, and rate constants may obtain a spatial dependence\cite{zwanzig1990rate}. With the advent of interfacial sensitive measurements\cite{somorjai1999molecular,kim2000ir,baldelli1998first,kott1993observation}, single molecule experiments\cite{austin1975dynamics,lu1998single,min2005fluctuating}, and precision electrochemical techniques\cite{garcia2010fluctuation,katemann2002fabrication}, quantifying how reactivity changes spatially at a molecular level is now possible. Theoretical work has trailed behind these advances, with few methods to efficiently study such processes and consequently few guiding principles for understanding how reactivity is altered in inhomogeneous systems. Here, we detail a theory and numerical technique to compute rate constants in the presence of spatial inhomogenieties without assuming that the mechanism of the transition is conserved at different points in space. This theory relies on the relationship between rate constants and trajectory space partition functions, and the method it motivates is general, capable of application to complex environments and processes. 

Rare dynamical events often take place in environments that are complex and inhomogeneous. Heterogeneous catalysis relies on the increase of the rate of a chemical reaction near an extended fluid-solid interface relative to the rate in the fluid\cite{somorjai2009reaction}. Moreover, heterogeneities along the interface like defects or grain boundaries can act as active sites for catalysis and have tremendous impact on reactivity\cite{weisz1970heterogeneous}. Reactivity can be influenced analogously by extended liquid-vapor interfaces, or by liquid-liquid interfaces\cite{benjamin1996chemical}, such as those that occur in atmospheric aerosols\cite{valsaraj2012review,bertram2009toward} or those that have been implicated in the rate enhancement of organic reactions using so-called `on-water' chemistry\cite{narayan2005water}. 

Beginning with pioneering work from Zwanzig\cite{zwanzig1990rate}, a theoretical formalism has been developed to understand the macroscopic implications of disorder on chemical kinetics, capable of describing the various limits of static and dynamic disorder. For a first order kinetic process occurring in an inhomogeneous system, whereby the concentration of some species, $A$, is changing, the differential rate expression is 
\begin{equation}
\frac{dA}{dt}= -k(y) A
\end{equation}
where the rate constant $k(y)$ depends in principle on where the reaction occurs, here parameterized by the coordinate $y$. The observed change in the concentration is given by the expectation value 
\begin{equation}
A(t) = A(0) \int dy\, p(y) e^{-k(y) t}
\end{equation}
where $p(y)$ is the equilibrium probability of $A$ being at $y$ at time 0. This expression makes an assumption that the disorder resulting in a functional dependence of $k$ on $y$ is static over timescales $t \approx 1/\mathrm{max}[k(y) ]$. This formalism is ubiquitous in the analysis of chemical kinetics and time-dependent spectroscopy. Despite this, few attempts have been made to study $k(y)$ directly theoretically or compute it numerically.

There are many computational techniques available to compute a rate constant\cite{peters2017reaction}. These can in principle be used to understand a rate constant's dependence on external static variables. However, doing so is often computationally cumbersome. Most commonly, a rate constant is computed by first intuiting a mechanistically relevant coordinate, computing the free energy along that coordinate and using the computed free energy barrier for a transition state theory estimate of the rate\cite{frenkel2001understanding}. Corrections to the transition state theory estimate can be evaluated using a Bennet-Chandler procedure, whereby the transmission coefficient is calculated and the exact rate constant established\cite{chandler1978statistical}. Alternatively, methods like forward flux sampling\cite{allen2006forward, allen2009forward} or transition interface sampling\cite{van2005elaborating, moroni2004investigating}, alleviate the need for positing a relevant reaction coordinate and compute a rate directly using a stratification strategy\cite{dinner2018trajectory}. In these and related techniques, the transition probability is estimated directly by breaking up the rare transition into a sequence of typical transitions. Both types of strategies could be extended to inhomogeneous systems, however to do so would require that these calculations be repeated at each representative area of space. While such calculations can in principle be done in parallel, their efficiency would rely on a conservation of the reactive mechanism. If the mechanism changed at different points in space, the adequacy of the order parameters introduced to compute the free energy or to stratify the transition probability would need to be changed.

Generically, a rate constant is computable within linear response theory from the time correlation function 
\begin{equation}
C(t) = \frac{\langle h_{A}(0)h_{B}(t)\rangle }{\langle h_{A}(0) \rangle}
\end{equation}
where $h_{A}$ and $h_{B}$ are indicator functions returning 1 if a configuration falls within the $A$ or $B$ basins, respectively, and 0 otherwise, and $\langle \dots \rangle$ denotes equilibrium average. For $\tau_\mathrm{mol} \ll t \ll 1/k_{{AB}}$  
\begin{equation}
C(t) \sim k_{AB} t
\end{equation}
where $\tau_\mathrm{mol}$ is a characteristic molecular relaxation time and $k_{{AB}}$ is the rate constant for transitioning between states ${A}$ and ${B}$.
This time correlation function can alternatively be viewed as a free energy in trajectory space\cite{dellago1999calculation}, or the ratio of constrained path partition functions.  If $P[\mathbf{r}(t)]$ is the probability of generating the trajectory, $\mathbf{r}(t)=\{r_0,r_1,\dots,r_t\}$ of length $t$, then $C(t)$ can be expressed as 
\begin{equation}
C(t) = \frac{\int \mathcal{D}[\mathbf{r}(t)]h_{A}[r_0]h_{B}[r_t]P[\mathbf{r}(t)]}{\int \mathcal{D}[\mathbf{r}(t)]h_{A}[r_0]P[\mathbf{r}(t)]} = \frac{\Omega_{AB}(t)}{\Omega_{A}}
\end{equation}
where the numerator, $\Omega_{AB}(t)$, is a path partition function for an ensemble of reactive trajectories, and the denominator, $\Omega_{A}$, is the equilibrium partition function constrained to beginning in the $A$ basin. 

We can extend this perspective to rate constants with spatial disorder by constraining the reactive path ensemble to visit a particular region of space. Specifically, we define a constrained path partition function,
\begin{eqnarray}
\Omega_{AB}(t , y) &=& \int \mathcal{D}[\mathbf{r}(t)]h_{A}[r_0]h_{B}[r_t]P[\mathbf{r}(t)]\delta(y-\hat{y}^i_{t^*})
\end{eqnarray}
where $\hat{y}^i_{t^*}$ is a spatial degree of freedom for the reactive species $i$ at the time $t^*$ commensurate with the commitment time of the reaction. Assuming that $y$ is slowly varying during the reaction, $\hat{y}_t \approx \hat{y}_{t*} \approx \hat{y}_0$, $y$ can be considered a static variable. The ratio of path partition functions constrained to different points in space, $y$ and $y'$, in the linearly reacting regime,
\begin{equation}
\label{Eq:Relative}
\frac{\Omega_{AB}(t, y)}{\Omega_{AB}(t,y')} = \frac{k_{AB}(y)}{k_{AB}(y')} e^{-\beta \Delta F}
\end{equation}
is equal to the relative rate for the reaction, weighted by the relative equilibrium probabilities of finding the reactants at $y$ relative to $y'$, given by the equilibrium free energy difference $\Delta F= F(y)-F(y')$ and $\beta$ is inverse temperature times Boltzmann's constant. 
This identification provides the opportunity for trajectory reweighing, in which if both $k_{AB}(y)$ and $\beta F(y)$ are known, the contribution to the rate from the stability of the reactants can be differentiated from the contribution from an increased barrier height or altered flux. Finally, if the absolute rates are desired one needs only to compute  $k_{AB}(y)$ at a single value of $y$. 

As with any ratio of partition functions, the ratio of path partition functions has an interpretation as a reversible work\cite{geissler2004equilibrium}, and therefore the rate extracted from its computation is independent of the specific pathway by which it is estimated. In principle Eq.~\ref{Eq:Relative} can be estimated by generating many reactive events and counting the number of reactions occurring at each value of $y$. Specifically, the path partition function conditioned at a specific value of $y$ is proportional to the probability of observing a reactive trajectory initiated at $y$, $P_{AB}(y)$, or consequently a ratio of path partition functions
\begin{equation}
\frac{P_{AB}(y)}{P_{AB}(y')}  = \frac{\Omega_{AB}(t, y)}{\Omega_{AB}(t,y')} 
\end{equation}
is equal to a ratio of conditioned probabilities. However, each reactive event is rare by definition, and moreover if the rates of these events are very different at different points in space, such brute force estimates are computationally intractable. The first problem of sampling rare reactive events can be solved using path sampling methods like transition path sampling (TPS)\cite{bolhuis2002transition}. In TPS, a procedure is designed to sample the ensemble of reactive trajectories. This can be accomplished by making random changes to an existing reactive trajectory and accepting the new trajectory with a Metropolis acceptance criteria, or by evolving the whole reactive trajectory collectively using the trajectory's action as an effective Hamiltonian\cite{dellago1998transition}. TPS and its extensions have been used for a broad range of applications, from studying chemical reactions\cite{geissler2001autoionization,quaytman2007reaction,saen2008atomic,knott2013mechanism} to biomolecular conformational changes \cite{hagan2003atomistic,juraszek2006sampling,juraszek2012transition}, to crystallization\cite{zahn2004atomistic,grunwald2009nucleation,beckham2011optimizing} and vitrification \cite{keys2011excitations,limmer2014theory} and even nonadiabatic dynamics\cite{schile2018studying}. Recently, TPS has been used to estimate path partition functions in order to evaluate transport coefficients\cite{gao2017transport} and large deviation functions out-of-equilibrium \cite{ray2018importance}. 

A given TPS calculation in an inhomogeneous system will not in general sample reactive events uniformly through space. Indeed since TPS is constructed to sample the reactive path ensemble in proportion to their statistical weight, Eq.~\ref{Eq:Relative} suggests that reactions will overwhelmingly occur in regions of space where the product of the rate times the Boltzmann weight is largest. Mapping out the relative rates as a function of some spatial coordinate would require sampling over rare fluctuations in trajectory space. This problem can be overcome just as it is in configurational Monte Carlo by adding a sampling bias to localize the path ensemble somewhere in space and correcting for that bias through histogram reweighting. In equilibrium this sort of procedure is known as umbrella sampling, and in that spirit we refer to the addition of a sample bias to TPS as TPS plus umbrella sampling or TPS+U. 

For detailed balanced dynamics, the two most common TPS moves are shooting and shifting. In each, provided a symmetry where the new trajectories are generated with the same dynamics as those of the desired path ensemble the acceptance criteria depends only on whether a newly generated trajectory is part of the conditioned reactive ensemble. 
In order to incorporate additional importance sampling into the TPS calculation, we bias the trajectory ensemble by tilting it exponentially 
\begin{equation}
\tilde{P}_{AB}(y) \propto P_{AB}(y) e^{-B(y)}
\end{equation}
where, $B(y)$, is an added sampling bias function, dependent on $y$, that can be used to localize reactive events to different positions in space. The added bias can be an arbitrary function of $y$, but in practice we will choose a quadratic form
$$
B(y) = \lambda (y-y^*)^2
$$
where $\lambda$ is the strength of the bias to localize around position $y^*$. In the limit that $\lambda$ is very large, probability will condense onto $y^*$ independent of the original distribution $P_{AB}(y)$, but in general $\tilde{P}_{AB}(y)$ reflects contributions from both factors. Using the same standard, symmetric shooting and shifting moves, this biased probability distribution can be sampled with a Metropolis acceptance criteria. Specifically, the acceptance criteria for making a random update to an old reference trajectory, $\mathbf{r}^\mathrm{o}(t)$, generating a new trajectory, $\mathbf{r}^\mathrm{n}(t)$, is
\begin{eqnarray}
\mathrm{Acc} [\mathbf{r}^\mathrm{o}(t)&& \rightarrow \mathbf{r}^\mathrm{n}(t)] =\\
&& \mathrm{min} \left [1,h_{A}(r_0^\mathrm{n}) h_{B}(r_t^\mathrm{n})e^{-B(y^\mathrm{n})+B(y^\mathrm{o}) } \right ] \nonumber
\end{eqnarray}
where $B(y^\mathrm{n})$ and $B(y^\mathrm{o})$ are the bias functions evaluated in the new and reference trajectories. 

Histogram reweighing techniques allow for the construction of the relative rate of the reactive process as a function of a degree of freedom that is static on the timescale of the reaction. 
This biased probability is related to the original reactive ensemble by
\begin{equation}
\ln P_{AB}(y) =\ln \tilde{P}_{AB}(y) + B(y) + \mathrm{Const.}
\end{equation}
where the additive constant is equivalent to the log ratio of the normalization constants between the biased and original ensemble. Often a set of biased ensembles with different values of $y^*$ is simulated, and provided sufficient overlap between the distributions, optimal reweighting techniques like the multistate Bennet acceptance ratio (MBAR)\cite{shirts2008statistically} or weighted histogram analysis method\cite{kumar1992weighted} can be used to determine the set of constants. 

\begin{figure}[t]
\begin{center}
\includegraphics[width=8.cm]{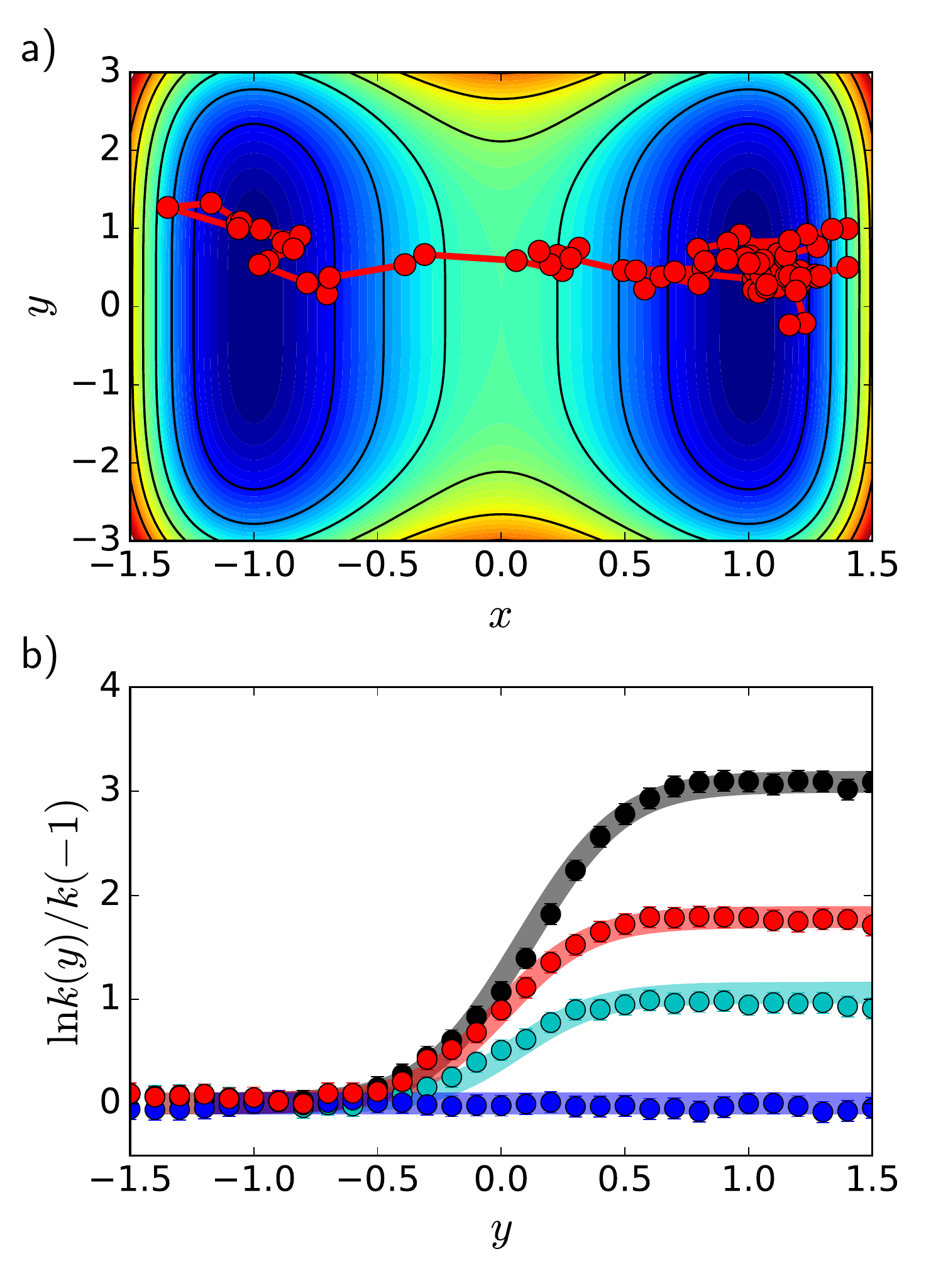}
\caption{Validation of the calculation of relative rate constants for a overdamped process with spatially varying diffusion. a) Two-dimensional potential, $U(x,y)$, with contour lines demarking every $1/\beta$ studied, and a characteristic reactive trajectory shown in red. b) Relative rate constant as a function of $y$, referenced at $y=-1$ for $\ln \Delta = 0$ (blue), 1 (cyan), 2 (red), and 3 (black), computed with TPS+U in symbols and evaluated exactly from Kramer's theory in solid lines. Errorbars are smaller than the symbols.}
\label{Fi:1}
\end{center} 
\end{figure}

To assess the accuracy of this methodology, we first study the overdamped motion of a particle in a two-dimensional potential with a spatially varying diffusivity. Specifically, we consider an equation of motion in two spatial dimensions ${r} = \{x,y\}$, with a spatially dependent diffusivity, $D({r})$, and potential $U({r})$,  
\begin{equation}
\dot{r} = -\beta D({r}) \nabla U({r}) + \eta(t)
 \end{equation}
where the dot denotes time derivative, $\eta$ is a Gaussian random variable, whose statistics are given by,
$\langle \eta(t) \rangle = 0$, and $\langle \eta(t)\eta(t') \rangle = 2 D({r}) \delta(t-t')$
and
\begin{equation}
\label{Eq:Pote}
U(x,y) = y^4/100-2x^2+x^4
\end{equation}
is a bimodal potential elongated in the $y$ direction. This potential is shown in Fig.~\ref{Fi:1}. For the spatially varying diffusivity we use,
\begin{equation}
D(x,y)/D_\mathrm{o} = \Delta/2 \left [ \tanh(5y)+1\right ]  + 1
\end{equation}
where $\Delta$ is the multiplicative change in the diffusivity for $y\ll 0$ relative to $y\gg 0$. This model has two symmetric stable minima with a single barrier in between. The size of the barrier is independent of the $y$, but the varying diffusivity results in rate for crossing the barrier that changes with $y$.  The line of saddle points in this model ensures that despite the anisotropic friction the particle must overcome the same barrier to transition\cite{berezhkovskii1989rate}. 

This is a convenient model to study as the rate for transitions between the state $x\approx -1$ and the state $x\approx 1$ as a function of $y$ can be computed exactly by solving the Smoluchowski equation, and the form of the relative rate constant in the $y$ direction is simply proportional to the spatially varying diffusivity\cite{zwanzig2001nonequilibrium}. As given by Kramer's theory, the rate from $A$ to $B$ for a diffusive system with spatially varying diffusion constant for fixed $y$ is given by
\begin{equation}
k_{AB}(y) = D(y) \left [ \int_{-\infty}^0 dx e^{-\beta U(x,y)}  \int_{-1}^{1} dx'  e^{\beta U(x',y)} \right ]^{-1}
\end{equation}
which is directly proportional to the spatially dependent diffusion constant. Due to the form of the potential in Eq.~\ref{Eq:Pote}, fluctuations in $x$ and $y$ are statistically uncorrelated, and the only dependence of the rate on $y$ will be through $D(y)$. 

To compute the $k_{AB}(y)$ numerically with TPS+U for this overdamped dynamics, we first define the indicator functions as
\begin{equation}
h_{A} = 1-\theta(x+1/2) \quad h_{B} = \theta(x-1/2)
\end{equation}
where $\theta(x)$ is a Heaviside step function. We choose parameters $\beta =4, D_\mathrm{o}=1$ and study the dependence on the offset in diffusion constant, $\Delta$. For this system we study a trajectory ensemble with $t=5$, which is within the linear growth regime for the side-side correlation function over the studied spatial range. In order to importance sample trajectories that occur at different values of  $y$, we simulate a series of windows with $\lambda=10$ and $y^*=\{-1.5:1.5\}$ in steps of 0.1 and we use MBAR to stitch the windows together. There is some choice in what specific value of $y$ to choose along the trajectory, since in practice reactions do not occur instantonically at one value of $y$. Ideally, the value of $y$ at the isocommitor surface\cite{vanden2010transition} should be used, but that in practice is difficult to determine. For reactions that have a strong separation of timescales between $\tau_\mathrm{mol}$ and $1/k_{AB}$ we have found that the specific choice does not qualitatively affect the results, and quantitatively changes the results only in regions where $k_{AB}(y)$ varies rapidly. In the following we use the value of $y$ at the midpoint in the trajectory. For each window, we perform 10000 combined one-sided shooting and shifting moves,\cite{brotzakis2016one} which is sufficient to have a well converged histogram.  

Figure \ref{Fi:1}b shows the calculation of the relative rate, $k_{AB}(y)/k_{AB}(y=-1)$ referenced to the rate at $y=-1$ for a set of increasingly larger diffusion constant differences, $\Delta$. This system is judiciously chosen such that over the range of $y=-1.5$ to $y=1.5$ the free energy changes little with $y$, so following from Eq.~\ref{Eq:Relative} the relative rate constant is given by the ratio of path space partition functions. For $\Delta =0$ the relative rate is flat as expected, and for increasing $\ln \Delta=1,2,$ and 3, we find sigmoidal curves that plateau to higher values. From Kramer's theory for an overdamped barrier crossing, the rate constant is proportional to $D(y)$ and indeed we find that the shapes for the relative rate constants follow exactly the form of the spatially dependent diffusion constant $D(y)$, which are plotted in solid curves on top of the measured data validating the sampling approach. 

In order to demonstrate the utility and feasibility of this method, we have also computed the relative rate constant for ion pair dissociation near and away from an electrochemical interface. In aqueous solution, the association or dissociation of oppositely charged ions is dynamically gated by collective rearrangement of surrounding water\cite{Geissler1999,Ballard2012,mullen2014transmission}. Electrostatic potentials generated through rare polarization fluctuations of the solvent are the rate determining step of dissociation, and these fluctuations can be dramatically altered in systems with extended inhomogeneities like that present at an electrode interface\cite{Limmer2013,willard2013characterizing}. Indeed previous work has shown that the dissociation of NaI ions near a water platinum interface can be slowed down by a factor of 40x relative to bulk water\cite{kattirtzi2017microscopic}. This is due to a higher free energy barrier to charge separation and a smaller flux over that barrier at the interface, resulting from strong water-electrode interactions that lead to the formation of an adsorbed water monolayer with preferential hydrogen bonding in-plane and concomitant slow orientational relaxation dynamics. With such a strong dependence of the rate of dissociation on the proximity to the interface, this system offers an ideal system for testing this method. While previous studies have noted that dissociation is accompanied by inertial effects at the top of the barrier,\cite{mullen2014transmission} we explicitly neglect these here and focus rather on the dependence of the rate arising from purely configurational contributions. 
\begin{figure}
\begin{center}
\includegraphics[width=8.5cm]{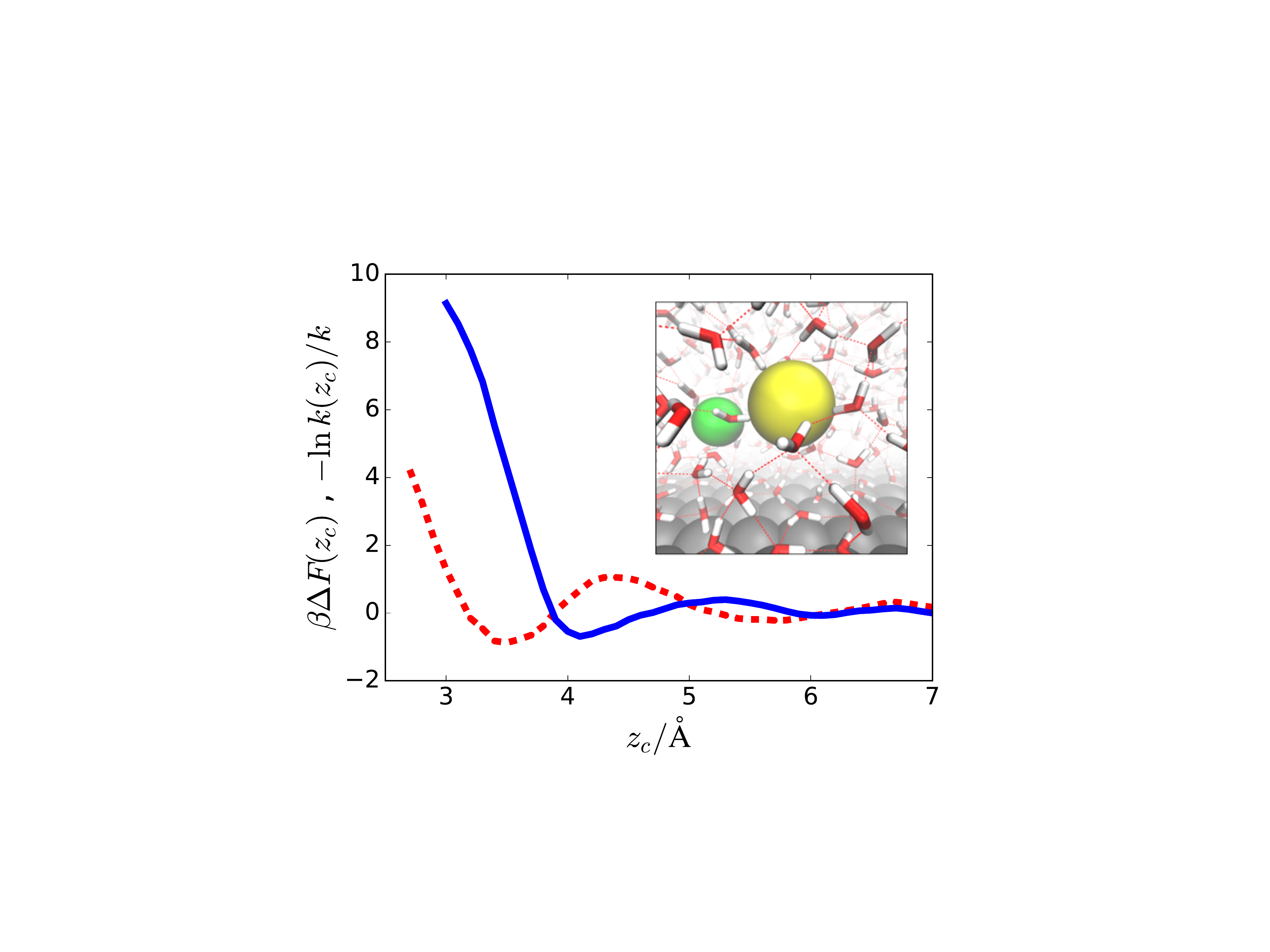}
\caption{Ion pair dissociation near an electrochemical interface. The free energy (dashed red) and relative rate constant (blue solid line) computed from TPS+U as a function of center of mass distance, $z_c$ from the electrochemical interface. (Inset) illustrates an the NaI near its transition state at $z_c=3.5 \mathrm{\AA}$.}
\label{Fi:3}
\end{center} 
\end{figure}

As was done in previous work,\cite{kattirtzi2017microscopic} we study a single NaI ion pair solvated in liquid water placed in contact with a planar 111 FCC surface of platinum. The water is modeled with the SPC/E potential\cite{berendsen1987missing} with geometry constrained using the SETTLE algorithm\cite{miyamoto1992settle}. The water electrode interaction is modeled with the potential proposed by Seipmann and Sprik\cite{siepmann1995influence}, and the ions interact with point charges and Lennard-Jones potentials\cite{koneshan1998solvent}, see Ref.~\onlinecite{kattirtzi2017microscopic} for details. The electrode is simulated at constant electrostatic potential\cite{siepmann1995influence,reed2007electrochemical,limmer2013charge}, and its atoms are held rigid. The surrounding solution is simulated with a constant number of particles, temperature, 300 K, and volume, using a Langevin thermostat with time constant 0.1 ps. A large friction is used in order to consider only the overdamped dynamics of the ion pair dissociation process and simplify analysis. Given the role of inertial effects found previously,\cite{mullen2014transmission} the large friction used here may limit the generality of our results. Nearly 1800 water molecules are used, in a simulation cell geometry of 3x3x4 nm. Simulations are performed in LAMMPS\cite{plimpton1995fast}. 

The relative rate calculation is performed on the component of center of mass of the ion pair perpendicular to the plane of the electrode, $z_c$. In order to calculate the relative equilibrium free energy along this coordinate, we employed umbrella sampling by adding a quadratic potential with spring constant 13.5 $1/\beta \mathrm{\AA}^2$ and 10 centers equally spaced between $z_c = \{2 : 9 \}\mathrm{\AA}$ exactly as in Ref.~\onlinecite{kattirtzi2017microscopic}. For the TPS+U calculations, we employed a reactive path ensemble defined by an observation time of 10 ps, and reactant and product basins defined by
\begin{equation}
h_{A} = 1-\theta(R-3.5 \mathrm{\AA}) \quad h_{B} = \theta(R-4.5 \mathrm{\AA})
\end{equation}
where $R$ is the inter-ion separation distance, which is taken to be 2 $1/\beta$ below the top of the free energy barrier computed. We used 15 windows with $\lambda=1$ and minima equally spaced along the ion center of mass, $z_c = \{2 : 9 \}\mathrm{\AA}$ at the midpoint along the trajectory. We employed 1500 single sided shooting and shifting moves with an average acceptance criterion of 0.3, resulting in approximately 500 uncorrelated trajectories for each window. MBAR was used to compute the relative path partition function, and we reference the rate at large $z_c$, denoted $k$.

Figure 3 shows the results of the TPS+U calculation, including the resulting relative rate constant and associated free energy along the distance from the water monolayer to the ion center of mass. The assumption of static disorder is satisfied due to the separation of timescales for the ions to diffuse relative to the surface and the instantonic timescale to dissociate. Both are referenced to their values at large $z_c$. The free energy is oscillatory, with a shallow minimum at contact with the adsorbed water monolayer, rises quickly for values of $z_c$ smaller that that minimum and plateaus to 0 at large values of $z_c$. This preferential surface adsorption is a consequence of the weak hydration of the ion pair.\cite{kattirtzi2017microscopic} The relative rate is constant at its bulk value and decreases dramatically at the interface. The sharpness of its fall reflects the highly localized boundary layer of water that within one layer transitions from predominately bulk like, to interfacially distinct.\cite{limmer2015nanoscale} The  decrease in the relative rate at the interface is in agreement with previous single point calculations that estimated the rate of dissociation at the interface at $z_c=3.5 \mathrm{\AA}$ and $z_c=8 \mathrm{\AA}$ using a flux-side correlation function. In that work, this decrease was attributed to an increased barrier height, and an additional decrease in the effective diffusion coefficient over the barrier.\cite{kattirtzi2017microscopic} To the extent that the the barrier is unchanged for small changes in $z_c$, the dependence of the rate corresponds to changes in the diffusion constant as the ions move in a medium of slower water molecules. This is confirmed by computing the diffusion constant along the Madelung potential, which was found to well describe the transition state ensemble.\cite{kattirtzi2017microscopic}

In this manuscript, we have provided a new perspective on dynamical events in inhomogenous systems by showing that the relative rate of a rare event is equivalent to a ratio of path space partition functions. Ratios of partition functions are equivalent to differences in free energies, which are typically much easier to compute than absolute free energies. Indeed we have shown that generalizing standard histogram reweighting techniques to path ensembles, the relative rate of a reactive event through space can be efficiently computed without prior mechanistic insight and free of bias. 

We have illustrated this perspective in a canonical complex reaction-- that of ion pair dissociation near an extended interface. We expect that other interfacial reactions can be studied using these methods analogously. More generally, rates of phase transformation in the presence of inhomogenieties, like heterogeneous nucleation, could be studied, or active sites in disordered catalytic systems could be uncovered with this machinery, provided only a computationally tractable representation of the system. Further decomposition of the ratio of path partition functions into energetic and flux contributions are possible through known path ensemble measures. For example, taking the derivative of the log ratio of the conditioned path partition functions with respect to $\beta$,
\begin{equation}
-\frac{d}{d \beta} \ln \frac{\Omega_{AB}(t, y)}{\Omega_{AB}(t,y')} = E^\dagger(y)-E^\dagger(y') + \Delta F
\end{equation}
we obtain the $y$ dependence on the activation barrier, $E^\dagger(y)$, which can be evaluated via a simple average within the conditioned path ensembles\cite{dellago2004activation, mesele2016removing}. Finally, generalizations beyond those discussed here could be envisioned. For example, rather than conditioning the reactive ensemble on different regions in space, we could condition on different product states in order to compute branching ratios without having to independently compute the rate of formation for each product. Such studies are on-going.

\begin{acknowledgments}
This material is based upon work supported by the U.S. Department of Energy, Office of Science, Office of Advanced Scientific Computing Research, Scientific Discovery through Advanced Computing (SciDAC) program under Award Number DE-AC02-05CH11231.  This research used resources of the National Energy Research Scientific Computing Center (NERSC), a U.S. Department of Energy Office of Science User Facility operated under Contract No. DE-AC02-05CH11231.
\end{acknowledgments}

\section*{References}
%

\end{document}